\documentclass[12pt]{article}
\usepackage{amsmath}

\oddsidemargin  - 5 mm
\evensidemargin - 5 mm

\input{tcilatex}

\begin{document}

\date{}
\title{Elliptic Curves and Algebraic Geometry Approach in Gravity Theory
III. Uniformization Functions for a Multivariable Cubic Algebraic Equation}
\author{Bogdan G. Dimitrov \thanks{%
Electronic mail: bogdan@theor.jinr.ru} \\
Bogoliubov Laboratory for Theoretical Physics\\
Joint Institute for Nuclear Research \\
6 Joliot - Curie str. \\
Dubna 141 980, Russia}
\maketitle

\begin{abstract}
\ \ The third part of the present paper continues the investigation of the
solution of the multivariable cubic algebraic equation for reparametrization
invariance of the gravitational Lagrangian. The main result in this paper
constitutes the fact that the earlier found parametrization functions of the
cubic algebraic equation for reparametrization invariance of the
gravitational Lagrangian can be considered also as uniformization functions.
These functions are obtained as solutions of first - order nonlinear
differential equations, as a result of which they depend only on the complex
(uniformization) variable $z$. Further, it has been demonstrated that this
uniformization can be extended to two complex variables, which is
particularly important for investigating various physical metrics, for
example the $ADS$ metric of constant negative curvature (Lobachevsky spaces).
\end{abstract}

\section{\protect\bigskip INTRODUCTION}

\bigskip In previous papers [1, 2] the general approach for investigation of
algebraic equations in gravity theory has been presented. The approach is
based essentially on the important distinction between covariant and
contravariant metric components in the framework of the \textit{%
gravitational theories with covariant and contravariant metrics and
connections (GTCCMC)}, which has been described in the review article [3].
In its essence, this distinction is related to the \textit{affine geometry
approach} [4,5], according to which the four-velocity tangent vector at each
point of the observer's worldline is \textit{not normalized} and equal to
one, i.e. $l_{a}l^{a}=l^{2}\neq 1$. Similarly, for a second-rank tensor one
would have $g_{\mu \nu }g^{\nu \alpha }=l_{\mu }^{\alpha }$ $\neq \delta
_{\mu }^{\alpha }$.

In [2] a solution for the chosen variables $dX^{1}$, $dX^{2}$, $dX^{3}$ in
terms of the elliptic Weierstrass function and its derivative has been found
of the earlier proposed [6] \textit{cubic algebraic equation of
reparametrization invariance of the gravitational Lagrangian} 
\begin{equation}
dX^{i}dX^{l}\left( p\Gamma _{il}^{r}g_{kr}dX^{k}-\Gamma
_{ik}^{r}g_{lr}d^{2}X^{k}-\Gamma _{l(i}^{r}g_{k)r}d^{2}X^{k}\right)
-dX^{i}dX^{l}R_{il}=0\text{ \ \ \ \ .}  \tag{1.1}
\end{equation}%
Finding the solution enables one to find the dependence of the contravariant
metric tensor components \ on the \textit{elliptic Weierstrass function and
its derivatives} and on variables, related to the covariant tensor and its
derivatives.

However, it is much more important to find the dependence of the covariant
tensor components. For some concrete cases - the Szafron-Szekeres
inhomogeneous cosmological model and its subcase - the FLRW (Friedman -
Lemaitre - Robertson - Walker) cosmology, such a representation has been
found in [7]. The general solution of inhomogeneous relativistic cosmology
was given in the form 
\begin{equation*}
ds^{2}=dt^{2}-\frac{\left( \frac{M(z)}{2}\right) ^{2}}{(\rho (u+\epsilon
)-\rho (v_{0}))^{2}}e^{2\nu }(dx^{2}+dy^{2})-
\end{equation*}%
\begin{equation}
-h^{2}(z)(\Phi ^{^{\prime }}+\Phi \nu ^{^{\prime }})^{2}dz^{2}\text{ \ \ .} 
\tag{1.2}
\end{equation}%
The function $\rho (v_{0})$ in [7] is supposed to satisfy the elliptic curve 
\begin{equation}
\left( \rho ^{^{\prime }}(v_{0})\right) ^{2}=4\rho ^{3}(v_{0})-g_{2}\rho
(v_{0})-g_{3}\text{ \ \ \ , }  \tag{1.3}
\end{equation}%
where $g_{2}$ and $g_{3}$ are assumed to be the functions 
\begin{equation}
g_{2}=\frac{K^{2}(z)}{12}\text{ \ \ ; \ \ \ }g_{3}=\frac{1}{216}K^{3}(z)-%
\frac{1}{12}\Lambda M^{2}(z)\text{ \ \ .}  \tag{1.7}
\end{equation}%
Unfortunately, the presented in [7] solution cannot be claimed to be true
due to the following simple reason: If $g_{2}$ and $g_{3}$ are functions and
at the same time, they satisfy the elliptic curve equation (1.3), then these
functions should be equal to the corresponding $\func{Ei}$\textit{senstein
series} (invariants) $g_{2}=60\sum\limits_{\omega \subset \Gamma }\frac{1}{%
\omega ^{4}}g_{3}=140\sum\limits_{\omega \subset \Gamma }\frac{1}{\omega ^{6}%
}$ . If this is taken into account, then it can easily be checked that the
obtained solution (1.2) would not be of that form - for example, there would
be no dependence on the $M(z)$ function in the second term.

Consequently, the problem about finding the correct solutions of the
Einstein's equations in terms of elliptic functions still remains open. But
even if solutions for the metric tensor are found in the form $g_{ij}(z,%
\mathbf{x)}$ (here $z$ is the complex variable, on which the Weierstrass
function depends, $\mathbf{x}$ are the coordinates in the chosen metric),
then this dependence on additional complex variable would still complicate
the solution.

In this third part of the paper, we shall consider an approach, when it will
be possible to find the solution not in terms of the \textit{parametrization
functions} $g_{ij}(z,\mathbf{x)}$, but in terms of \textit{uniformization
functions} - these are the functions, which depend only on the complex
variable $z$ (or, as we shall see, on two complex variables $z$ and $v$) and 
\textit{on no other variables}. In fact, it will be shown that after solving
a \textit{first - order system of nonlinear differential equations} with
respect to the generalized coordinates $X^{i}$, their dependence on the
complex coordinate $z$ can be found. Thus, an important problem from the
point of algebraic geometry will be solved - the uniformization functions $%
dX^{i}=dX^{i}(z)$ for the multivariable cubic algebraic equation (1.1) are
found, as a result also the metric tensor $g_{ij}(z,X(z))=g_{ij}(z)$. \ \ \
\ 

\section{\protect\bigskip\ \ COMPLEX \ COORDINATE \ DEPENDENCE \ OF \ THE \
METRIC \ TENSOR \ COMPONENTS \ FROM \ THE \ UNIFORMIZATION \ OF \ A \ CUBIC
\ ALGEBRAIC \ SURFACE}

\bigskip In this section it will be shown that the solutions (2. 16), (2.
21) and (2. 28) (see [2] - Part II) of the cubic algebraic equation (1.1)
enable us to express not only the contravariant metric tensor components
through the Weierstrass function and its derivatives, but the covariant
components as well.

Let us write down for convenience the system of equations (2. 16), (2. 21)
and (2. 28) for $dX^{1}$, $dX^{2}$ and $dX^{3}$ as ($l=1,2,3$) 
\begin{equation}
dX^{l}(X^{1},X^{2},X^{3})=F_{l}(g_{ij}(\mathbf{X}),\Gamma _{ij}^{k}(\mathbf{X%
}),\rho (z),\rho ^{^{\prime }}(z))=F_{l}(\mathbf{X},z)\text{ \ \ \ ,} 
\tag{2.1}
\end{equation}%
\ where the appearence of the complex coordinate $z$ is a natural
consequence of the uniformization procedure, applied with respect to each
one of the cubic equations from the \textquotedblright
embedded\textquotedblright\ sequence of equations.

Yet how the appearence of the additional complex coordinate $z$ on the R. H.
S. of (2.1) can be reconciled with the dependence of the differentials on
the L. H. S. only on the generalized coordinates $(X^{1},X^{2},X^{3})$ (and
on the initial coordinates $x^{1},x^{2},x^{3}$ because of the mapping $%
X^{i}=X^{i}(x^{1},x^{2},x^{3})$)? The only reasonable assumption will be
that \textit{the initial coordinates depend also on the complex coordinate},
i.e. 
\begin{equation}
X^{l}\equiv X^{l}(x^{1}(z),x^{2}(z),x^{3}(z))=X^{l}(\mathbf{x,}\text{ }z)%
\text{ \ \ \ \ .}  \tag{2.2}
\end{equation}

\bigskip\ Taking into account the important initial assumptions ($l=1,2,3$) 
\begin{equation}
d^{2}X^{l}=0=dF_{l}(\mathbf{X}(z),z)=\frac{dF_{l}}{dz}dz\text{ \ \ ,} 
\tag{2.3}
\end{equation}%
one easily gets the system of three inhomogeneous linear algebraic equations
with respect to the functions $\frac{\partial X^{1}}{\partial z}$, $\frac{%
\partial X^{2}}{\partial z}$ and $\frac{\partial X^{3}}{\partial z}$ ($%
l=1,2,3$): 
\begin{equation}
\frac{\partial F_{l}}{\partial X^{1}}\frac{\partial X^{1}}{\partial z}+\frac{%
\partial F_{l}}{\partial X^{2}}\frac{\partial X^{2}}{\partial z}+\frac{%
\partial F_{l}}{\partial X^{3}}\frac{\partial X^{3}}{\partial z}+\frac{%
\partial F_{l}}{\partial z}=0\text{ \ \ \ ,}  \tag{2.4}
\end{equation}%
The solution of this algebraic system ($i,k,l=1,2,3$) 
\begin{equation}
\frac{\partial X^{l}}{\partial z}=G_{l}\left( \frac{\partial F_{i}}{\partial
X^{k}}\right) =G_{l}\left( X^{1},X^{2},X^{3},z\right) \text{ \ \ \ \ \ } 
\tag{2.5}
\end{equation}%
represents a system of \textit{three first - order nonlinear differential
equations}.\textbf{\ }A solution of this system can always be found in the
form \textbf{\ } 
\begin{equation}
X^{1}=X^{1}(z)\text{ \ \ ; \ \ \ }X^{2}=X^{2}(z)\text{ \ \ ; \ \ \ \ }%
X^{3}=X^{3}(z)\text{ \ \ \ \ \ \ \ \ \ .}  \tag{2.6}
\end{equation}%
and therefore, the metric tensor components will also depend on the complex
coordinate $z$, i.e. $g_{ij}=g_{ij}(\mathbf{X}(z))$. Note that since the
functions $\frac{\partial F_{i}}{\partial X^{k}}$ in the R. H. S. of (2.5)
depend on the Weierstrass function and its derivatives, it might seem
natural to write that \textbf{\ }the solution of the above system of
nonlinear differential equations $g_{ij}$\ will also depend on the
Weierstrass function and its derivatives 
\begin{equation}
g_{ij}=g_{ij}(X^{1}(\rho (z),\rho ^{^{\prime }}(z),X^{2}(\rho (z),\rho
^{^{\prime }}(z),X^{3}(\rho (z),\rho ^{^{\prime }}(z))=g_{ij}(z)\text{ \ \ \
.}  \tag{2.7}
\end{equation}%
Note however that for the moment we do not have a theorem that the solution
of the system (2.5) will also contain the Weierstrass function\textbf{. }But
the dependence on the complex coordinate $z$ will be retained.

Now let us mention the other equations, which will further be taken into
account.

The first set of equations simply means that the differentials $dF_{1}$, $%
dF_{2}$, $dF_{3}$, equal to the second differentials $d^{2}X^{1}$, $%
d^{2}X^{2}$, $d^{2}X^{3}$ can be taken with respect both to the generalized
coordinates $X^{1}$, $X^{2}$, $X^{3}$ and the initial coordinates $x^{1}$, $%
x^{2}$, $x^{3}$ ($l=1,2,3$)%
\begin{equation}
d^{2}X^{l}=dF_{l}(\mathbf{X}(z),z)=dF_{l}(\mathbf{x}(z),z)\text{ \ \ \ \ .} 
\tag{2.8}
\end{equation}%
\ Denoting further $\overset{.}{x}^{1}\equiv \frac{\partial x^{1}}{\partial z%
}$, $\overset{.}{x}^{2}\equiv \frac{\partial x^{2}}{\partial z}$ and $%
\overset{.}{x}^{3}\equiv \frac{\partial x^{3}}{\partial z}$, the above
equalities result again in a system of three inhomogeneous algebraic
equations with respect \ to $\overset{.}{X}^{1}\equiv \frac{\partial X^{1}}{%
\partial z}$, $\overset{.}{X}^{2}\equiv \frac{\partial X^{2}}{\partial z}$
and $\overset{.}{X}^{3}\equiv \frac{\partial X^{3}}{\partial z}$ 
\begin{equation}
\frac{\partial F_{l}}{\partial X^{1}}\frac{\partial X^{1}}{\partial z}+\frac{%
\partial F_{l}}{\partial X^{2}}\frac{\partial X^{2}}{\partial z}+\frac{%
\partial F_{l}}{\partial X^{3}}\frac{\partial X^{3}}{\partial z}=\frac{%
\partial F_{l}}{\partial x^{1}}\overset{.}{x}^{1}+\frac{\partial F_{l}}{%
\partial x^{2}}\overset{.}{x}^{2}+\frac{\partial F_{l}}{\partial x^{3}}%
\overset{.}{x}^{3}\text{.}  \tag{2.9}
\end{equation}%
Assuming for the moment that we know the functions $\overset{.}{x}^{1}$, $%
\overset{.}{x}^{2}$ and $\overset{.}{x}^{3}$, the solutions of this
algebraic system will give again another system of three first - order
nonlinear differential equations ($l=1,2,3$) 
\begin{equation}
\frac{\partial X^{l}}{\partial z}=H_{l}\left( X^{1},X^{2},X^{3},\text{ }z%
\text{ },\text{ }\overset{.}{x}^{1},\overset{.}{x}^{2},\overset{.}{x}%
^{3}\right) \text{ \ \ \ \ \ .}  \tag{2.10}
\end{equation}%
Again, a solution of this system like the one in (2.6) can be obtained but
with account of the dependence on the additional variables $\overset{.}{x}%
^{1}$, $\overset{.}{x}^{2}$ and $\overset{.}{x}^{3}$. Let us also here note
that the solution (2.6) of the nonlinear system of equations (2. 5) can be
assumed to be dependent on some another complex variable $v$ 
\begin{equation}
X^{1}=X^{1}(z,v)\text{ \ \ \ ; \ \ }X^{1}=X^{1}(z,v)\text{ \ \ \ ; \ \ \ }%
X^{1}=X^{1}(z,v)\text{\ \ \ \ \ .}  \tag{2.11}
\end{equation}%
The system of equations (2.8) ($i=1,2,3$) 
\begin{equation}
d^{2}X^{i}=dF_{i}(\mathbf{X}(z,v),z)=dF_{1}(\mathbf{x}(z,v),z)\text{ \ \ \
,\ }  \tag{2.12}
\end{equation}%
with account of the expressions (2.10) now will be rewritten as 
\begin{equation*}
\frac{\partial F_{i}}{\partial X^{1}}\frac{\partial X^{1}}{\partial v}+\frac{%
\partial F_{i}}{\partial X^{2}}\frac{\partial X^{2}}{\partial v}+\frac{%
\partial F_{i}}{\partial X^{3}}\frac{\partial X^{3}}{\partial v}=\frac{%
\partial F_{i}}{\partial x^{1}}\overset{.}{x}^{1}+\frac{\partial F_{i}}{%
\partial x^{2}}\overset{.}{x}^{2}+\frac{\partial F_{i}}{\partial x^{3}}%
\overset{.}{x}^{3}+
\end{equation*}%
\ \ \ 
\begin{equation}
+\frac{\partial F_{i}}{\partial x^{1}}x^{^{\prime }1}+\frac{\partial F_{i}}{%
\partial x^{2}}x^{^{\prime }2}+\frac{\partial F_{i}}{\partial x^{3}}%
x^{^{\prime }3}-\frac{\partial F_{i}}{\partial X^{1}}H_{1}-\frac{\partial
F_{i}}{\partial X^{2}}H_{2}-\frac{\partial F_{i}}{\partial X^{3}}H_{3}=0%
\text{ \ ,}  \tag{2.13}
\end{equation}%
where $x^{^{\prime }1},x^{^{\prime }2},x^{^{\prime }3}$ denote the
derivatives $\frac{\partial x^{1}}{\partial z},\frac{\partial x^{2}}{%
\partial z},\frac{\partial x^{3}}{\partial z}$. The same notation further
will be used with respect to the variables $\frac{\partial X^{1}}{\partial v}%
,\frac{\partial X^{2}}{\partial v},\frac{\partial X^{3}}{\partial v}$ .
Similarly to (2. 10), the algebraic solution of this system of equations can
be represented as 
\begin{equation}
\frac{\partial X^{i}}{\partial v}=K_{i}\left( \mathbf{X(}z,v\mathbf{),}\text{
}z\text{ },\overset{.}{\mathbf{x}}\text{ },\mathbf{x}^{^{\prime }}\right) 
\text{ \ \ \ .}  \tag{2.14}
\end{equation}%
Note that instead of (2.12), we could have also written 
\begin{equation}
d^{2}X^{i}=dF_{i}(\mathbf{X}(z,v),z)=dF_{1}(\mathbf{x}(z),z,v)\text{ \ \ \
.\ }  \tag{2.15}
\end{equation}

\bigskip Further in section 3 it shall be proved why this would be
incorrect. The complete analysis of the system of equations, when both
system of coordinates depend on the two pair of complex variables $z$ and $v$
will be given in the following sections. For the moment we give just the
general qualitative motivations.

The other set of equations, which will further be used and which relates the
generalized coordinates $X^{i}$ to the initial ones $x^{i}$ is 
\begin{equation}
d^{2}X^{i}=0=\frac{\partial ^{2}X^{i}}{\partial x^{k}\partial x^{r}}%
dx^{k}dx^{r}+\frac{\partial X^{i}}{\partial x^{k}}d^{2}x^{k}\text{ \ \ .} 
\tag{2.16}
\end{equation}%
For the moment we assume that the initial coordinates $x^{k}$ depend only on
the $z$ coordinate, and therefore 
\begin{equation}
\frac{\partial ^{2}X^{i}}{\partial x^{k}\partial x^{r}}=\frac{\overset{..}{X}%
^{i}}{\overset{.}{x}^{k}\overset{.}{x}^{r}}-\overset{.}{X}^{i}\frac{\overset{%
..}{x}^{r}}{\overset{.}{x}^{k}\left( \overset{.}{x}^{r}\right) ^{2}}\text{ \
\ .}  \tag{2.17}
\end{equation}

Taking this into account, the system (2.16) in the $n-$dimensional case can
be written as 
\begin{equation}
n^{2}\overset{..}{X}^{i}(dz)^{2}-(n-1)\overset{.}{X}^{i}\frac{\overset{..}{x}%
^{r}}{\overset{.}{x}^{r}}(dz)^{2}+n\overset{.}{X}^{i}d^{2}z=0\text{ \ \ \ .}
\tag{2.18}
\end{equation}%
Introducing the notation 
\begin{equation}
y^{r}=\frac{\partial }{\partial z}\left( \ln \overset{.}{x}^{r}\right) =%
\frac{\overset{..}{x}^{r}}{\overset{.}{x}^{r}}\text{ \ \ \ \ \ }  \tag{2.19}
\end{equation}%
for the three-dimensional case, the system (2.18) can be written as 
\begin{equation}
2\overset{.}{X}^{i}(dz)^{2}(y^{1}+y^{2}+y^{3})=9\overset{..}{X}^{i}(dz)^{2}+3%
\overset{.}{X}^{i}d^{2}z\text{ \ \ \ \ .}  \tag{2.20}
\end{equation}%
\bigskip\ Dividing the L. H. S. and the R. H. S. of the $i$-th and the $j$%
-th equation of this system, it can easily be obtained 
\begin{equation}
\left( dz\right) ^{2}\left( \overset{..}{X}^{i}\overset{.}{X}^{j}-\overset{..%
}{X}^{j}\overset{.}{X}^{i}\right) =0\text{ \ \ \ \ ,}  \tag{2.21}
\end{equation}%
which can be written as 
\begin{equation}
(dz)^{2}\left( \overset{.}{X}^{j}\right) ^{2}\frac{\partial }{\partial z}%
\left( \frac{\overset{.}{X}^{i}}{\overset{.}{X}^{j}}\right) =0\text{ \ \ \ \
.}  \tag{2.22}
\end{equation}%
\ Neglecting the case when $\overset{.}{X}^{j}=0$, the above relation simply
means that $\overset{.}{X}^{2}$ and $\overset{.}{X}^{3}$should be
proportional to $\overset{.}{X}^{1}$%
\begin{equation}
\overset{.}{X}^{2}=C_{2}\overset{.}{X}^{1}\text{ \ ; \ }\overset{.}{X}%
^{3}=C_{3}\overset{.}{X}^{1}\text{ \ \ ,}  \tag{2.23}
\end{equation}%
where $C_{2}$ and $C_{3}$ are constants. Indeed, it is easily seen that
(2.23) holds since 
\begin{equation}
dX^{1}=\overset{.}{X}^{1}dz=F_{1}\text{ ; \ \ \ }dX^{2}=\overset{.}{X}%
^{2}dz=F_{2}\text{ ; \ \ \ }dX^{3}=\overset{.}{X}^{3}dz=F_{3}  \tag{2.24}
\end{equation}%
and consequently 
\begin{equation}
C_{2}=\frac{F_{2}}{F_{1}}\text{ \ \ \ \ ; \ \ \ \ }C_{3}=\frac{F_{3}}{F_{1}}%
\text{ \ \ \ .}  \tag{2.25}
\end{equation}

\section{\protect\bigskip FIRST-ORDER \ NONLINEAR \ DIFFERENTIAL \ EQUATIONS
\ FOR \ THE \ COMPLEX \ FUNCTIONS \ $x=x(z)$ AND \ $X=X(z)$}

\bigskip For the purpose, the two systems of algebraic equations (2.4) and
(2.9) will be used. If one substitutes the found expressions (2.23) for $%
\overset{.}{X}^{2}$and \ $\overset{.}{X}^{3}$into the system (2.4), it may
be treated as an algebraic system of equations with respect to the variables 
$\overset{.}{X}^{1},C_{2}$ and $C_{3}$. Introducing the notation 
\begin{equation}
\{F_{i},F_{j}\}_{z,X^{k}}\equiv \frac{\partial F_{i}}{\partial z}\frac{%
\partial F_{j}}{\partial X^{k}}-\frac{\partial F_{i}}{\partial X^{k}}\frac{%
\partial F_{j}}{\partial z}\text{ \ \ }  \tag{3.1}
\end{equation}%
for the \textquotedblright one-dimensional\textquotedblright\ \textit{%
Poisson bracket} $\{F_{i},F_{j}\}_{z,X^{k}}$ of the coordinates $z,X^{k}$
and also the notation 
\begin{equation}
\{F_{1},F_{2},F_{3}\}_{z,\left[ X^{i},X^{j}\right] }\equiv
\{F_{1},F_{2}\}_{z,X^{i}}\{F_{1},F_{3}\}_{z,X^{j}}-\{F_{1},F_{2}\}_{z,X^{j}}%
\{F_{1},F_{3}\}_{z,X^{i}}\text{ \ \ \ ,}  \tag{3.2}
\end{equation}%
one can show that the solution of the system of \textit{linear algebraic
equations} (2.4) with respect to $\overset{.}{X}^{1},C_{2}$ and $C_{3}$ 
\begin{equation}
\frac{\partial F_{i}}{\partial X^{1}}\overset{.}{X}^{1}+\frac{\partial F_{i}%
}{\partial X^{2}}\overset{.}{X}^{2}+\frac{\partial F_{i}}{\partial X^{3}}%
\overset{.}{X}^{3}+\frac{\partial F_{i}}{\partial z}=0  \tag{3.3}
\end{equation}%
can be represented in the following compact form 
\begin{equation}
C_{2}=\frac{\{F_{1},F_{2},F_{3}\}_{z,\left[ X^{3},X^{1}\right] }}{%
\{F_{1},F_{2},F_{3}\}_{z,\left[ X^{2},X^{3}\right] }}\text{ \ \ \ ; \ \ \ \
\ }C_{3}=\frac{\{F_{1},F_{2},F_{3}\}_{z,\left[ X^{1},X^{2}\right] }}{%
\{F_{1},F_{2},F_{3}\}_{z,\left[ X^{2},X^{3}\right] }}\text{ \ \ \ ,\ } 
\tag{3.4}
\end{equation}%
\begin{equation}
\overset{.}{X}^{1}=-\frac{\frac{\partial F_{1}}{\partial z}%
\{F_{1},F_{2},F_{3}\}_{z,\left[ X^{2},X^{3}\right] }}{K_{1}}\text{ \ \ \ \ .}
\tag{3.5}
\end{equation}%
In (3.5) the following notation has been introduced for $K_{i}$ ($i=1,2,3$) 
\begin{equation}
K_{i}\equiv \frac{\partial F_{i}}{\partial X^{1}}\{F_{1},F_{2},F_{3}\}_{z,%
\left[ X^{2},X^{3}\right] }+\frac{\partial F_{i}}{\partial X^{2}}%
\{F_{1},F_{2},F_{3}\}_{z,\left[ X^{3},X^{1}\right] }+\frac{\partial F_{i}}{%
\partial X^{3}}\{F_{1},F_{2},F_{3}\}_{z,\left[ X^{1},X^{2}\right] }\text{ \
\ \ \ .}  \tag{3.6}
\end{equation}%
The usefulness of introducing this notation will soon be understood.

Now let us rewrite the system of equations (2.9) in the form 
\begin{equation}
\frac{\partial F_{i}}{\partial x^{1}}\overset{.}{x}^{1}+\frac{\partial F_{i}%
}{\partial x^{2}}\overset{.}{x}^{2}+\frac{\partial F_{i}}{\partial x^{3}}%
\overset{.}{x}^{3}=M_{i}\text{ \ \ \ \ ,}  \tag{3.7}
\end{equation}%
where $M_{i}$ will be the notation for 
\begin{equation}
M_{i}\equiv \frac{\partial F_{i}}{\partial X^{1}}\overset{.}{X}^{1}+\frac{%
\partial F_{i}}{\partial X^{2}}\overset{.}{X}^{2}+\frac{\partial F_{i}}{%
\partial X^{3}}\overset{.}{X}^{3}\text{ \ \ .}  \tag{3.8}
\end{equation}%
Making use of the above formulaes (3.4 - 3.6) and also (2.23), $M_{i}$ can
be calculated to be 
\begin{equation}
M_{i}=-\frac{\partial F_{1}}{\partial z}\frac{K_{i}}{K_{1}}\text{ \ \ \ .} 
\tag{3.9}
\end{equation}%
Further, the solutions of the linear algebraic system of equations (3.7) can
be represented in the form 
\begin{equation}
\overset{.}{x}^{i}=S_{1}^{i}M_{1}+S_{2}^{i}M_{2}+S_{3}^{i}M_{3}\text{ \ \ \
\ ,}  \tag{3.10}
\end{equation}%
where the functions $S_{1}^{i},S_{2}^{i}$ and $S_{3}^{i}$ depend on $\frac{%
\partial F_{i}}{\partial x^{k}}$ ($i,k=1,2,3$). Since $M_{1}$, $M_{2}$, $%
M_{3}$ according to (3.9) and (3.6) are proportional to $\frac{\partial F_{1}%
}{\partial z}\frac{\{F_{1},F_{2},F_{3}\}_{z,\left[ X^{k},X^{j}\right] }}{%
K_{1}}$ (where $(k,j)=(2,3)$, $(3,1)$ or $(1,2)$), the resulting solution
(3.10) will be of the kind 
\begin{equation}
\overset{.}{x}^{i}=\frac{\overline{S}_{1}^{i}\left( \frac{\partial F_{1}}{%
\partial z}\right) ^{2}+\overline{S}_{2}^{i}\frac{\partial F_{2}}{\partial z}%
\frac{\partial F_{1}}{\partial z}+\overline{S}_{3}^{i}\frac{\partial F_{1}}{%
\partial z}\frac{\partial F_{3}}{\partial z}}{\overline{S}_{4}^{i}\left( 
\frac{\partial F_{1}}{\partial z}\right) +\overline{S}_{5}^{i}\frac{\partial
F_{2}}{\partial z}+\overline{S}_{6}^{i}\frac{\partial F_{3}}{\partial z}}%
\text{ \ \ \ \ ,}  \tag{3.11}
\end{equation}

where the functions $\overline{S}_{1}^{i},\overline{S}_{2}^{i},....,%
\overline{S}_{6}^{i}$ depend both on $\frac{\partial F_{i}}{\partial X^{k}}$
and $\frac{\partial F_{i}}{\partial x^{k}}$ and consequently on all the
variables $x^{k},X^{k}$ and $z$. We have used also the following relation,
obtained after simple algebra with account of (3.1) and (3.2) 
\begin{equation*}
\{F_{1},F_{2},F_{3}\}_{z,\left[ X^{i},X^{j}\right] }=\left( \frac{\partial
F_{1}}{\partial z}\right) ^{2}\{F_{2},F_{3}\}_{X^{i},X^{j}}+
\end{equation*}%
\begin{equation}
+\frac{\partial F_{1}}{\partial z}\frac{\partial F_{2}}{\partial z}%
\{F_{3},F_{1}\}_{X^{i},X^{j}}+\frac{\partial F_{1}}{\partial z}\frac{%
\partial F_{3}}{\partial z}\{F_{1},F_{2}\}_{X^{i},X^{j}}\text{ \ \ .} 
\tag{3.12}
\end{equation}%
Thus we have obtained the system of \textit{first order nonlinear
differential equations} with  respect to the initial coordinates $%
x^{i}=x^{i}(z)$\textbf{.} An analogous system of nonlinear differential
equations is obtained for $X^{1}=X^{1}(z)$, $X^{2}=X^{2}(z)$ and $%
X^{3}=X^{3}(z)$ - for $X^{1}$ this is equation (3.5), and with account of
(2.23) and expressions (3.4) for $C_{2}$ and $C_{3}$, the corresponding
equations for $X^{2}(z)$ and $X^{3}(z)$ are 
\begin{equation}
\overset{.}{X}^{2}=-\frac{\frac{\partial F_{1}}{\partial z}%
\{F_{1},F_{2},F_{3}\}_{z,\left[ X^{3},X^{1}\right] }}{K_{1}}\text{ \ \ ; \ }%
\overset{.}{X}^{3}=-\frac{\frac{\partial F_{1}}{\partial z}%
\{F_{1},F_{2},F_{3}\}_{z,\left[ X^{1},X^{2}\right] }}{K_{1}}\text{.} 
\tag{3.13}
\end{equation}%
Therefore, if the generalized coordinates $X^{1},X^{2},X^{3}$ are determined
as functions of the complex variable $z$ after solving the system (3.5),
(3.13), the obtained functions $X^{1}=X^{1}(z)$, $X^{2}=X^{2}(z)$ and $%
X^{3}=X^{3}(z)$ can be substituted into the R. H.\ S. of the system (3.11)
for $x^{1},x^{2}$ and $x^{3}$ and the corresponding solutions $x^{1}=x^{1}(z)
$, $x^{2}=x^{2}(z)$ and $x^{3}=x^{3}(z)$ can be found. Remember that we
started from the assumption that only the generalized coordinates $%
X^{1},X^{2},X^{3}$ satisfy the original cubic algebraic equation and
therefore equalities (2.24) are fulfilled. Nevertheless, the corresponding
functions $x^{i}=x^{i}(z)$\ is possible to be determined from the system
(3.11), the R. H. S. of which also confirms that $dx^{i}\neq F_{i}$.

This conclusion is important since it shows that the two systems of
coordinates should not be treated on an equal footing. This refers of course
to the case of only one complex coordinate.

\section{\protect\bigskip IS IT \ POSSIBLE \ TO \ HAVE \ A \ TWO \ COMPLEX \
COORDINATE \ DEPENDENCE \ OF \ THE \ GENERALIZED \ COORDINATES \ $%
X^{i}=X^{i}\left( \mathbf{x(}z\right) ,z,v)?$ \ }

It will be proved below that such a case should be disregarded since it
leads to an impossibility to determine the dependence $X^{i}$ on the $v$
coordinate.

Under the above assumption $X^{i}=X^{i}\left( \mathbf{x(}z\right) ,z,v)$,
the first set of three equations 
\begin{equation}
dX^{i}=\frac{\partial X^{i}}{\partial x^{1}}dx^{1}+\frac{\partial X^{i}}{%
\partial x^{2}}dx^{2}+\frac{\partial X^{i}}{\partial x^{3}}dx^{3}\text{ \ \ }
\tag{4.1}
\end{equation}%
can be represented as 
\begin{equation}
F_{i}=\overset{.}{X}^{i}\frac{\partial z}{\partial x^{1}}\overset{.}{x}%
^{1}dz+\overset{.}{X}^{i}\frac{\partial z}{\partial x^{2}}\overset{.}{x}%
^{2}dz+\overset{.}{X}^{i}\frac{\partial z}{\partial x^{3}}\overset{.}{x}%
^{3}dz=3\overset{.}{X}^{i}dz\text{ \ \ \ ,}  \tag{4.2}
\end{equation}%
so again relations (2.23) - (2.25) $\ \overset{.}{X}^{2}=\frac{F_{2}}{F_{1}}%
\overset{.}{X}^{1}$, \ $\overset{.}{X}^{3}=\frac{F_{3}}{F_{1}}\overset{.}{X}%
^{1}$will hold.

The second set of equations 
\begin{equation}
d^{2}X^{i}=dF_{i}(X,z)=dF_{i}(X(z,v),z)=dF_{i}(z,v)\text{ \ \ \ \ } 
\tag{4.3}
\end{equation}%
will express the equality of the differentials, expressed in terms of the
two different sets of coordinates $(X,z)$ and $(z,v)$ 
\begin{equation*}
d^{2}X^{i}=\frac{\partial F_{i}}{\partial X^{1}}dX^{1}+\frac{\partial F_{i}}{%
\partial X^{2}}dX^{2}+\frac{\partial F_{i}}{\partial X^{3}}dX^{3}+\frac{%
\partial F_{i}}{\partial z}dz=
\end{equation*}%
\begin{equation*}
=\left[ \frac{\partial F_{i}}{\partial X^{1}}\overset{.}{X}^{1}+\frac{%
\partial F_{i}}{\partial X^{2}}\overset{.}{X}^{2}+\frac{\partial F_{i}}{%
\partial X^{3}}\overset{.}{X}^{3}+\frac{\partial F_{i}}{\partial z}\right]
dz+
\end{equation*}%
\begin{equation}
+\left[ \frac{\partial F_{i}}{\partial X^{1}}dX^{^{\prime }1}+\frac{\partial
F_{i}}{\partial X^{2}}dX^{^{\prime }2}+\frac{\partial F_{i}}{\partial X^{3}}%
dX^{^{\prime }3}\right] dv\text{ \ \ \ .}  \tag{4.4}
\end{equation}%
Taking into account that according to (2.25) $dX_{1}=F_{1}$, $dX_{2}=F_{2}$
and $dX_{3}=F_{3}$ and also the expressed from (2.25) differential 
\begin{equation}
dz=\frac{1}{3}\frac{F_{1}}{\overset{.}{X}^{1}}\text{ \ \ ,}  \tag{4.5}
\end{equation}%
one can obtain for (4.4) 
\begin{equation*}
\left[ \frac{\partial F_{i}}{\partial X^{1}}dX^{^{\prime }1}+\frac{\partial
F_{i}}{\partial X^{2}}dX^{^{\prime }2}+\frac{\partial F_{i}}{\partial X^{3}}%
dX^{^{\prime }3}\right] dv=
\end{equation*}%
\begin{equation}
=\frac{2}{3}\left[ \frac{\partial F_{i}}{\partial X^{1}}F_{1}+\frac{\partial
F_{i}}{\partial X^{2}}F_{2}+\frac{\partial F_{i}}{\partial X^{3}}F_{3}\right]
\text{ \ \ \ \ \ .}  \tag{4.6}
\end{equation}%
Dividing the L. H. S. and the R. H. S. for different values of the indice $%
i=1,2,3$, one can obtain the following system of linear homogeneous
algebraic equations with respect to $X^{^{\prime }1},X^{^{\prime }2}$ and $%
X^{^{\prime }3}$ (the indice $i$ takes values $1,2,3,1,2..,$i.e. if $i=3\,$,
then $i+1$ would be $1$) 
\begin{equation*}
\left( \frac{\partial F_{i}}{\partial X^{1}}Q_{i+1}-\frac{\partial F_{i+1}}{%
\partial X^{1}}Q_{i}\right) X^{^{\prime }1}+\left( \frac{\partial F_{i}}{%
\partial X^{2}}Q_{i+1}-\frac{\partial F_{i+1}}{\partial X^{2}}Q_{i}\right)
X^{^{\prime }2}+
\end{equation*}%
\begin{equation}
+\left( \frac{\partial F_{i}}{\partial X^{3}}Q_{i+1}-\frac{\partial F_{i+1}}{%
\partial X^{3}}Q_{i}\right) X^{^{\prime }3}=0\text{ \ \ ,}  \tag{4.7}
\end{equation}%
where $Q_{i}$ ($i=1,2,3$) denotes the expression 
\begin{equation}
Q_{i}\equiv \frac{\partial F_{i}}{\partial X_{1}}F_{1}+\frac{\partial F_{i}}{%
\partial X^{2}}F_{2}+\frac{\partial F_{i}}{\partial X^{3}}F_{3}\text{ \ \ .}
\tag{4.8}
\end{equation}%
Note that for the moment we have not yet used the equations $%
d^{2}X^{i}=dF_{i}=0$, from where $Q_{i}=0$. Then the system of equations
(4.7) would be identically satisfied for all $X^{^{\prime }1},X^{^{\prime }2}
$ and $X^{^{\prime }3}$ and it would be impossible to express them as
solutions of the system. But even without making use of the equations $%
d^{2}X^{i}=dF_{i}=0$, the consistency (or inconsistency) of the system (4.7)
is a necessary condition for the consistency (or inconsistency) of the
assumption about $X^{i}=X^{i}\left( \mathbf{x(}z\right) ,z,v)$.

Making use of the notation (3.1), the determinant of the system can be
written as 
\begin{equation}
\begin{vmatrix}
\dsum\limits_{l_{1}\neq 1}F_{l_{1}}\{F_{1},F_{2}\}_{1,l_{1}} & 
\dsum\limits_{l_{2}\neq 2}F_{l_{2}}\{F_{1},F_{2}\}_{2,l_{2}} & 
\dsum\limits_{l_{3}\neq 3}F_{l_{3}}\{F_{1},F_{3}\}_{3,l_{3}} \\ 
\dsum\limits_{m_{1}\neq 1}F_{m_{1}}\{F_{1},F_{3}\}_{1,m_{1}} & 
\dsum\limits_{m_{2}\neq 2}F_{m_{2}}\{F_{1},F_{3}\}_{2,m_{2}} & 
\dsum\limits_{m_{3}\neq 3}F_{m_{3}}\{F_{1},F_{3}\}_{3,m_{3}} \\ 
\dsum\limits_{n_{1}\neq 1}F_{n_{1}}\{F_{2},F_{3}\}_{1,n_{1}} & 
\dsum\limits_{n_{2}\neq 2}F_{n_{2}}\{F_{2},F_{3}\}_{2,n_{2}} & 
\dsum\limits_{n_{3}\neq 3}F_{n_{3}}\{F_{2},F_{3}\}_{3,n_{3}}%
\end{vmatrix}%
\text{ \ \ \ \ ,}  \tag{4.9}
\end{equation}%
where instead of $\{F_{i},F_{j}\}_{X^{k},X^{n_{k}}}$ we have written only $%
\{F_{i},F_{j}\}_{k,n_{k}}$ and each element in the determinant represents a
sum either over $l_{1}$, $l_{2}$ or $l_{3}$.

The explicite calculation of the determinant (4.9) gives the non-zero
expression 
\begin{equation}
\frac{\partial F_{2}}{\partial X^{1}}\{F_{1},F_{3}\}_{2,3}+\frac{\partial
F_{3}}{\partial X^{1}}\{F_{1},F_{2}\}_{1,2}\text{ \ \ .}  \tag{4.10}
\end{equation}%
Since the determinant is non-zero, the system of linear homogeneous
algebraic equations does not have a solution and consequently the assumption
that $X^{i}=X^{i}\left( \mathbf{x(}z\right) ,z,v)$ turns out to be incorrect.

\section{\protect\bigskip COMPLEX \ STRUCTURE \ \ $X^{i}=X^{i}\left( \mathbf{%
x(}z,v\right) ,z) $ $\ $OF \ \ THE \ GENERALIZED \ COORDINATES \ AND \ OF \
THE \ METRIC \ TENSOR \ COMPONENTS}

\textbf{\bigskip }Now it shall be proved that the parametrization (2.1) of
the initially given cubic algebraic curve (surface) can be extended to a
parametrization in terms of a pair of \ complex coordinates $(z,v)$ and thus
a \textit{complex structure} can be introduced. Of particular interest in
view of possible physical applications to theories with extra dimensions and
relation to $ADS$ \ theories, which will be discussed in the conclusion,
will be the case of $v=\overline{z}$, when a pair of holomorphic -
antiholomorphic variables can be introduced.

In principle a manifold may admit a complex structure [8], if it can be
covered with opened sets $U$ $_{1},V_{1},U_{2},V_{2}.....$, such that in any
intersection $U_{i}\cap V_{i}$ the associated transformations $z^{k^{\prime
}}=z^{k^{\prime }}(z_{i},v_{i})$ are complex (analytical) functions.

The investigated problem may be formulated as follows. Let (again) the
system of equations (2.1) is given, subjected to the additional constraining
equation $d^{2}X^{i}=0$. Then the parametrization (2.1) of the initially
given cubic algebraic surface can \ be extended to a parametrization by
means of a pair of complex coordinates $(z,v)$\ in the following way 
\begin{equation}
dX^{i}(\mathbf{X})=F_{i}(\mathbf{X}(\mathbf{x}(z,v)),z)\text{ \ \ .} 
\tag{5.1}
\end{equation}%
Therefore, it should be proved that the same system of equations,
investigated in the previous sections, is not contradictable under the
assumption $X^{i}=X^{i}\left( \mathbf{x(}z,v\right) ,z)$.

The \textit{first set of equations} to be used is similar to (4.3), but this
time expressing the equality of the differentials 
\begin{equation}
dF_{i}(\mathbf{X}(z,v),z)\text{ }=dF_{i}(\mathbf{x}(z,v),z)\text{ \ \ \ \ ,}
\tag{5.2}
\end{equation}%
written in terms of the coordinates $(\mathbf{X,}z)$ and $(\mathbf{x},z)$ 
\begin{equation*}
\left[ \frac{\partial F_{i}}{\partial X^{1}}\overset{.}{X}^{1}+\frac{%
\partial F_{i}}{\partial X^{2}}\overset{.}{X}^{2}+\frac{\partial F_{i}}{%
\partial X^{3}}\overset{.}{X}^{3}+\frac{\partial F_{i}}{\partial z}\right]
dz+
\end{equation*}%
\begin{equation*}
+\left[ \frac{\partial F_{i}}{\partial X^{1}}X^{^{\prime }1}+\frac{\partial
F_{i}}{\partial X^{2}}X^{^{\prime }2}+\frac{\partial F_{i}}{\partial X^{3}}%
X^{^{\prime }3}\right] dv=
\end{equation*}%
\begin{equation*}
=\left[ \frac{\partial F_{i}}{\partial x^{1}}\overset{.}{x}^{1}+\frac{%
\partial F_{i}}{\partial x^{2}}\overset{.}{x}^{2}+\frac{\partial F_{i}}{%
\partial x^{3}}\overset{.}{x}^{3}+\frac{\partial F_{i}}{\partial z}\right]
dz+
\end{equation*}%
\begin{equation}
+\left[ \frac{\partial F_{i}}{\partial x^{1}}x^{^{\prime }1}+\frac{\partial
F_{i}}{\partial x^{2}}x^{^{\prime }2}+\frac{\partial F_{i}}{\partial x^{3}}%
x^{^{\prime }3}\right] dv\text{ \ \ \ \ .}  \tag{5.3}
\end{equation}%
The \textit{second set of equations} takes into account the fact that the
second differential $d^{2}X^{i}$ is zero, or equivalently 
\begin{equation}
d^{2}X^{i}=dF_{i}(\mathbf{X}(z,v),z)=0\text{ \ \ \ \ \ ,}  \tag{5.4}
\end{equation}%
where $dF_{i}(\mathbf{X}(z,v),z)$ is given by the L. H. S. of equation (5.3).

The \textit{third set of equations} is 
\begin{equation}
dX^{i}=F^{i}=\frac{\partial X_{i}}{\partial z}dz+\frac{\partial X_{i}}{%
\partial v}dv\text{ \ \ \ \ .}  \tag{5.5}
\end{equation}%
Let us now introduce the notations 
\begin{equation}
M_{i}(X,z)\equiv \frac{\partial F_{i}}{\partial X^{k}}\overset{.}{X}^{k}%
\text{ \ \ ; \ \ \ \ }M_{i}(x,z)\equiv \frac{\partial F_{i}}{\partial x^{k}}%
\overset{.}{x}^{k}\text{ \ \ ,}  \tag{5.6}
\end{equation}%
\begin{equation}
M_{i}(X,v)\equiv \frac{\partial F_{i}}{\partial X^{k}}X^{^{\prime }k}\text{
\ \ ; \ \ \ \ }M_{i}(x,v)\equiv \frac{\partial F_{i}}{\partial x^{k}}%
x^{^{\prime }k}\text{ \ \ ,}  \tag{5.7}
\end{equation}%
which will allow us to write down the  first and the second set of equations
(4.3) - (4.4) in the following compact form 
\begin{equation}
\left[ M_{i}(X,z)-M_{i}(x,z)\right] dz+\left[ M_{i}(X,v)-M_{i}(x,v)\right]
dv=0\text{ \ \ ,}  \tag{5.8}
\end{equation}%
\ 
\begin{equation}
\left[ M_{i}(X,z)+\frac{\partial F_{i}}{\partial z}\right] dz+M_{i}(X,v)dv=0%
\text{ \ \ .}  \tag{5.9}
\end{equation}%
Expressing $\frac{\partial X^{i}}{\partial v}dv$ from (5.5), it can easily
be proved that 
\begin{equation}
M_{i}(X,v)dv=\frac{\partial F_{i}}{\partial X^{k}}X^{^{\prime }k}dv=-\frac{%
\partial F_{i}}{\partial X^{k}}\overset{.}{X}^{k}dz+\frac{\partial F_{i}}{%
\partial X^{k}}F_{k}\text{ \ \ \ \ ,}  \tag{5.10}
\end{equation}%
where the last term is zero due to the fulfillment of the second set of
equations (5.5). Consequently, from (5.10) it follows 
\begin{equation}
M_{i}(X,z)dz+M_{i}(X,v)dv=dF_{i}(z,v)=dF_{i}(\mathbf{X}(z,v),z)=0\text{ \ \
\ .}  \tag{5.11}
\end{equation}%
Additionally, if (5.11) is substracted from (5.8) and (5.9), one easily
obtains 
\begin{equation}
M_{i}(x,z)dz+M_{i}(x,v)dv=dF_{i}(z,v)=dF_{i}(\mathbf{x}(z,v),z)=0\text{ \ \
\ ,}  \tag{5.12}
\end{equation}%
\begin{equation}
\frac{\partial F_{i}}{\partial z}dz=0\text{ \ \ \ }\Rightarrow \text{ \ \ }%
\frac{\partial F_{i}}{\partial v}dv=0\text{ \ \ \ \ .}  \tag{5.13}
\end{equation}%
In other words, if the differential $dF_{i}(\mathbf{X}(z,v),z)$ is zero in
terms of the coordinates $(\mathbf{X,}z\mathbf{)}$, then it necessarily
should be zero in the coordinates $(\mathbf{x},z)$. But in the spirit of the
discussion at the end of section 2, this does not mean that if $dX^{i}=F_{i}$%
, then the same should hold also for the initial coordinates $x^{i}$, i.e. $%
dx^{i}=F_{i}$. Indeed, we can find 
\begin{equation*}
M_{i}(x,z)\equiv \frac{\partial F_{i}}{\partial x^{k}}\overset{.}{x}^{k}=%
\frac{\partial F_{i}}{\partial X^{l}}\frac{\partial X^{l}}{\partial x^{k}}%
\overset{.}{x}^{k}=
\end{equation*}%
\begin{equation}
=\frac{\partial F_{i}}{\partial X^{l}}\left[ \frac{\overset{.}{X}^{l}}{%
\overset{.}{x}^{k}}+\frac{X^{^{\prime }l}}{x^{^{\prime }k}}\right] \overset{.%
}{x}^{k}=M_{i}(X,z)+M_{i}(X,v)\frac{\overset{.}{x}^{k}}{x^{^{\prime }k}}%
\text{ \ \ \ \ .}  \tag{5.14}
\end{equation}%
Similarly 
\begin{equation}
M_{i}(x,v)=M_{i}(X,v)+M_{i}(X,z)\frac{x^{^{\prime }k}}{\overset{.}{x}^{k}}%
\text{ \ \ .}  \tag{5.15}
\end{equation}%
If the above two expressions are substituted into (5.12), and (5.11) is
taken into account, one can obtain 
\begin{equation}
M_{i}(X,v)\frac{\overset{.}{x}^{k}}{x^{^{\prime }k}}dz+M_{i}(X,z)\frac{%
x^{^{\prime }k}}{\overset{.}{x}^{k}}dv=0\text{ \ \ \ .}  \tag{5.16}
\end{equation}%
Additionally, we have 
\begin{equation}
M_{i}(X,v)=M_{i}(X,z)\frac{x^{^{\prime }k}}{\overset{.}{x}^{k}}\text{ \ \ \
; \ \ \ \ \ }dv=-\frac{M_{i}(x,z)}{M_{i}(x,v)}dz\text{ \ \ \ \ .}  \tag{5.17}
\end{equation}%
Therefore, the following equation in partial derivatives with respect to $%
F_{i}=F_{i}(x^{l})$ can be derived 
\begin{equation}
\frac{\partial F_{i}}{\partial x^{l}}x^{^{\prime }l}\frac{\overset{.}{x}^{k}%
}{x^{^{\prime }k}}\frac{x^{^{\prime }m}}{\overset{.}{x}^{m}}-\frac{\partial
F_{i}}{\partial x^{l}}\overset{.}{x}^{l}\frac{x^{^{\prime }k}}{\overset{.}{x}%
^{k}}=0\text{ \ \ \ \ .}  \tag{5.18}
\end{equation}%
A stronger statement may be proved, clearly showing that from $%
dF_{i}(X,z)=dF_{i}(x,z)=0$ and $dX_{i}=F_{i}$ it does not follow that $%
dx^{i}=F_{i}$. If expression (4.17) for $M_{i}(X,v)=M_{i}(X,z)\frac{%
x^{^{\prime }k}}{\overset{.}{x}^{k}}$ is substituted into (5.11), one
obtains 
\begin{equation}
M_{i}(X,z)\left[ dz+\frac{x^{^{\prime }k}}{\overset{.}{x}^{k}}dv\right] =0 
\tag{5.22}
\end{equation}%
and since $M_{i}(X,z)\neq 0$ (and if $\frac{\partial F_{i}}{\partial X^{l}}%
\neq 0$), it follows 
\begin{equation}
dx^{k}=\overset{.}{x}^{k}dz+x^{^{\prime }k}dv=0\text{ \ \ \ .}  \tag{5.23}
\end{equation}

\section{\protect\bigskip ANALYSIS \ OF \ THE \ FOURTH \ AND \ THE \ FIFTH \
SET \ OF \ EQUATIONS \ FOR \ THE \ PREVIOUS \ CASE \ $X^{i}=X^{i}(\mathbf{x}%
(z,v),z)$}

\bigskip The \textit{fourth set of equations}, which will be considered is 
\begin{equation*}
dX^{k}=\frac{\partial X^{k}}{\partial x^{1}}dx^{1}+\frac{\partial X^{k}}{%
\partial x^{2}}dx^{2}+\frac{\partial X^{k}}{\partial x^{3}}dx^{3}=
\end{equation*}%
\begin{equation}
=3\overset{.}{X}^{k}dz+X^{^{\prime }k}\frac{\overset{.}{x}^{m}}{x^{^{\prime
}m}}dz+\overset{.}{X}^{k}\frac{x^{^{\prime }m}}{\overset{.}{x}^{m}}%
dv+X^{^{\prime }k}dv\text{ \ \ \ \ .}  \tag{6.1}
\end{equation}%
If multiplied by $\frac{\partial F_{i}}{\partial X^{k}}dzdv$ and also
relation (5.11) $M_{i}(X,z)dz+M_{i}(X,v)dv=0$ is taken into account, the
fourth set of equations can be written as 
\begin{equation}
\frac{\partial F_{i}}{\partial X^{k}}F_{k}dv=M_{i}(X,z)dz\left[ \frac{%
x^{^{\prime }m}}{\overset{.}{x}^{m}}\left( dv\right) ^{2}-\frac{\overset{.}{x%
}^{m}}{x^{^{\prime }m}}\left( dz\right) ^{2}\right] \text{ \ \ \ \ .} 
\tag{6.2}
\end{equation}%
The \textit{fifth set} of equations is 
\begin{equation}
d^{2}X^{k}=0=\frac{\partial ^{2}X^{k}}{\partial x^{m}\partial x^{n}}%
dx^{m}dx^{n}+\frac{\partial X^{k}}{\partial x^{m}}d^{2}x^{m}\text{ \ \ ,} 
\tag{6.3}
\end{equation}%
where the expressions on the R.H.S. can easily be computed in terms of the
coordinates $X$ and $x$ and their derivatives.

Our goal further will be to see \textit{whether the fifth equation (6.3)
constitutes a separate equation or whether it follows from the preceeding
four ones.}

For the purpose, let us multiply both sides of the fifth equation by $\frac{%
\partial F_{i}}{\partial X^{k}}$ and see which are the terms, containing the
second differentials $d^{2}z$ and $d^{2}v$ 
\begin{equation*}
\left( \frac{\partial F_{i}}{\partial X^{k}}\frac{\overset{.}{X}^{k}}{%
\overset{.}{x}^{m}}+\frac{\partial F_{i}}{\partial X^{k}}\frac{X^{^{\prime
}k}}{x^{^{\prime }m}}\right) \left( \overset{.}{x}^{m}d^{2}z+x^{^{\prime
}m}d^{2}v\right) =
\end{equation*}%
\begin{equation}
=M_{i}(x,z)d^{2}z+M_{i}(x,v)d^{2}v\text{ \ \ .}  \tag{6.4 }
\end{equation}%
In (6.4) we have used relations (5.14) and (5.15) for $M_{i}(x,z)$ and $%
M_{i}(x,v)$. But we may note that the obtained term in (6.4) can be found
from the relation (5.12) $M_{i}(x,z)dz+M_{i}(x,v)dv=0$, if it is
differentiated by $z$ and $v$ and the resulting equations are summed up.
Therefore 
\begin{equation*}
M_{i}(x,z)d^{2}z+M_{i}(x,v)d^{2}v=-M_{i}(x,z)(dz)^{2}-M_{i}(x,v)(dv)^{2}-
\end{equation*}%
\begin{equation}
-(\overset{.}{M}_{i}(x,v)+M_{i}^{^{\prime }}(x,z))dxdv\text{ \ \ \ . } 
\tag{6.5 }
\end{equation}%
The derivatives $\overset{.}{M}_{i}(x,v)$ and $M_{i}^{^{\prime }}(x,z)$ can
be found also from the already used expressions (5.14) and (5.15) 
\begin{equation*}
M_{i}^{^{\prime }}(x,z)=M_{i}^{^{\prime }}(X,z)+\frac{\overset{.}{x}^{m}}{%
x^{^{\prime }m}}M_{i}^{^{\prime }}(X,v)+
\end{equation*}%
\begin{equation}
+M_{i}(X,v)\frac{\left[ \overset{.}{x}^{^{\prime }m}x^{^{\prime }m}-\overset{%
.}{x}^{m}x^{^{\prime \prime }m}\right] }{\left( x^{^{\prime }m}\right) ^{2}}%
\text{ \ \ \ \ ,}  \tag{6.6 }
\end{equation}%
\begin{equation*}
\overset{.}{M}_{i}(x,v)=\overset{.}{M}_{i}(X,v)+\frac{x^{^{\prime }m}}{%
\overset{.}{x}^{m}}\overset{.}{M}_{i}(X,z)+
\end{equation*}%
\begin{equation}
+M_{i}(X,z)\frac{\left[ \overset{.}{x}^{^{\prime }m}\overset{.}{x}%
^{m}-x^{^{\prime }m}\overset{.}{x}^{m}\right] }{\left( x^{^{\prime
}m}\right) ^{2}}\text{ \ \ \ \ .}  \tag{6.7}
\end{equation}

\bigskip Making use of all the expressions (6.4) - (6.7), the following
expression for the fifth equation (6.3), multiplied by $\frac{\partial F_{i}%
}{\partial X^{k}}$, can be obtained: 
\begin{equation*}
(dz)^{2}[-2M_{i}(X,z)+M_{i}(X,z)\frac{\overset{..}{x}^{m}}{\overset{.}{x}^{m}%
}+\frac{\partial F_{i}}{\partial X^{k}}\frac{\partial ^{2}X^{k}}{\partial
x^{m}\partial x^{n}}\overset{.}{x}^{m}\overset{.}{x}^{n}]+
\end{equation*}%
\begin{equation*}
+(dv)^{2}[-2\frac{x^{^{\prime }m}}{\overset{.}{x}^{m}}M_{i}(X,z)+M_{i}(X,z)%
\frac{x^{^{\prime \prime }m}}{\overset{.}{x}^{m}}+\frac{\partial F_{i}}{%
\partial X^{k}}\frac{\partial ^{2}X^{k}}{\partial x^{m}\partial x^{n}}%
x^{^{\prime }m}x^{^{\prime }n}]+
\end{equation*}%
\begin{equation*}
+dzdv[2M_{i}(X,z)\frac{\overset{.}{x}^{^{\prime }m}}{\overset{.}{x}^{m}}+2%
\frac{\partial F_{i}}{\partial X^{k}}\frac{\partial ^{2}X^{k}}{\partial
x^{m}\partial x^{n}}\overset{.}{x}^{m}x^{^{\prime }n}-2M_{i}^{^{\prime
}}(X,z)-
\end{equation*}%
\begin{equation}
-2\overset{.}{M_{i}}(X,z)\frac{x^{^{\prime }m}}{\overset{.}{x}^{m}}%
+2M_{i}(X,z)\frac{\left[ x^{^{\prime }m}\overset{..}{x}^{m}-\overset{.}{x}%
^{^{\prime }m}\overset{.}{x}^{m}\right] }{\left( \overset{.}{x}^{m}\right)
^{2}}]=0\text{ \ \ .}  \tag{6.8 }
\end{equation}%
The last two terms $\frac{\partial F_{i}}{\partial X^{k}}\frac{\partial
^{2}X^{k}}{\partial x^{m}\partial x^{n}}\overset{.}{x}^{m}\overset{.}{x}^{n}$%
and $\frac{\partial F_{i}}{\partial X^{k}}\frac{\partial ^{2}X^{k}}{\partial
x^{m}\partial x^{n}}x^{^{\prime }m}x^{^{\prime }n}$ in the first two square
brackets can be found as follows: First, the derivatives $\overset{.}{M}%
_{i}(X,v)$ and $M_{i}^{^{\prime }}(X,v)$ can be expressed from the relation
(4.17) $M_{i}(X,v)=M_{i}(X,z)\frac{x^{^{\prime }k}}{\overset{.}{x}^{k}}$: 
\begin{equation}
\overset{.}{M_{i}}(X,v)=\overset{.}{M_{i}}(X,z)\frac{x^{^{\prime }m}}{%
\overset{.}{x}^{m}}+M_{i}(X,z)\frac{\left[ \overset{.}{x}^{^{\prime }m}%
\overset{.}{x}^{m}-x^{^{\prime }m}\overset{..}{x}^{m}\right] }{\left( 
\overset{.}{x}^{m}\right) ^{2}}\text{ \ \ \ ,}  \tag{6.9 }
\end{equation}%
\begin{equation}
M_{i}^{^{\prime }}(X,v)=M_{i}^{^{\prime }}(X,z)\frac{x^{^{\prime }m}}{%
\overset{.}{x}^{m}}+M_{i}(X,z)\frac{\left[ x^{^{\prime \prime }m}\overset{.}{%
x}^{m}-x^{^{\prime }m}\overset{.}{x}^{^{\prime }m}\right] }{\left( \overset{.%
}{x}^{m}\right) ^{2}}\text{ \ \ \ .}  \tag{6.10 }
\end{equation}%
But on the other hand, the same derivatives can be found by using the
defining expressions (5.6 - 5.7) 
\begin{equation*}
\overset{.}{M}_{i}(X,v)=\frac{\partial ^{2}F_{i}}{\partial X^{k}\partial
X^{l}}\frac{\partial X^{l}}{\partial x^{m}}\frac{\partial X^{k}}{\partial
x^{n}}\overset{.}{x}^{m}x^{^{\prime }n}+\frac{\partial F_{i}}{\partial X^{k}}%
\frac{\partial ^{2}X^{k}}{\partial x^{m}\partial x^{n}}\overset{.}{x}%
^{n}x^{^{\prime }m}+
\end{equation*}%
\begin{equation}
+M_{i}(X,z)\frac{x^{^{\prime }m}\overset{.}{x}^{n}}{\overset{.}{x}^{m}}%
+M_{i}(X,z)\overset{.}{x}^{^{\prime }n}\text{ \ \ ,}  \tag{6.11 }
\end{equation}%
\begin{equation*}
M_{i}^{^{\prime }}(X,v)=\frac{\partial ^{2}F_{i}}{\partial X^{k}\partial
X^{l}}\frac{\partial X^{l}}{\partial x^{m}}\frac{\partial X^{k}}{\partial
x^{n}}x^{^{\prime }m}x^{^{\prime }n}+\frac{\partial F_{i}}{\partial X^{k}}%
\frac{\partial ^{2}X^{k}}{\partial x^{m}\partial x^{n}}x^{^{\prime
}n}x^{^{\prime }m}+
\end{equation*}%
\begin{equation}
+\frac{\partial F_{i}}{\partial X^{k}}\frac{\partial X^{k}}{\partial x^{n}}%
\frac{x^{^{\prime }m}\overset{.}{x}^{^{\prime }n}}{\overset{.}{x}^{m}}+\frac{%
\partial F_{i}}{\partial X^{k}}\frac{\partial X^{k}}{\partial x^{m}}%
x^{^{\prime \prime }n}\text{ \ \ .}  \tag{6.12 }
\end{equation}%
Therefore, the desired expressions can be found by setting up formulae (6.9)
equal to (6.11) and also formulae (6.10) equal to (6.12). It can easily be
derived how eq. (6.8) will transform, but unfortunately, this would not
result in any simplification of the equation with respect to the generalized
coordinates $X^{i}$.

It remained only to show whether the fourth equation (5.2) is independent
from the preceeding ones (and thus can be treated separately) or it follows
naturally from these equations. For the purpose, let us take the
differential of (5.2) and use equations (5.10). After some lengthy, but
straightforward calculations it can be obtained 
\begin{equation*}
(dz)(dv)^{2}[\overset{.}{M_{i}}(X,z)\frac{x^{^{\prime }m}}{\overset{.}{x}^{m}%
}+2\frac{x^{^{\prime }m}}{\overset{.}{x}^{m}}M_{i}^{^{\prime
}}(X,z)-M_{i}(X,z)\frac{\left( \overset{.}{x}^{^{\prime }m}\overset{.}{x}%
^{m}-x^{^{\prime }m}\overset{..}{x}^{m}\right) }{\left( \overset{.}{x}%
^{m}\right) ^{2}}+
\end{equation*}%
\begin{equation*}
+M_{i}(X,z)\frac{2\left( \overset{.}{x}^{^{\prime }m}\overset{.}{x}%
^{m}-x^{^{\prime }m}\overset{..}{x}^{m}+x^{^{\prime \prime }m}\overset{.}{%
\overset{.}{x}^{m}-\overset{.}{x}^{^{\prime }m}x^{^{\prime }m}}\right) }{%
\left( \overset{.}{x}^{m}\right) ^{2}}]+
\end{equation*}%
\begin{equation*}
+(dz)^{2}(dv)[M_{i}^{^{\prime }}(X,z)\frac{\overset{.}{x}^{m}}{x^{^{\prime
}m}}+2\overset{.}{M}_{i}(X,z)+M_{i}(X,z)\frac{\left( \overset{.}{x}%
^{^{\prime }m}x^{^{\prime }m}-\overset{.}{x}^{m}x^{^{\prime \prime
}m}\right) }{\left( x^{^{\prime }m}\right) ^{2}}]+
\end{equation*}%
\begin{equation*}
+(dz)^{3}[\overset{.}{M_{i}}(X,z)\frac{\overset{.}{x}^{m}}{x^{^{\prime }m}}%
+M_{i}(X,z)\frac{\left( \overset{..}{x}^{m}x^{^{\prime }m}-\overset{.}{x}^{m}%
\overset{.}{x}^{^{\prime }m}\right) }{\left( x^{^{\prime }m}\right) ^{2}}]+
\end{equation*}%
\begin{equation}
+(dv)^{3}[M_{i}^{^{\prime }}(X,z)\frac{x^{^{\prime }m}}{\overset{.}{x}^{m}}%
+M_{i}(X,z)\frac{\left( x^{^{\prime \prime }m}\overset{.}{x}^{m}-x^{^{\prime
}m}\overset{.}{x}^{^{\prime }m}\right) }{\left( \overset{.}{x}^{m}\right)
^{2}}]=0\text{ \ \ \ \ \ .}  \tag{6.13 }
\end{equation}%
This is an equation both for the initial coordinates $x^{i}$ and for the
generalized ones $X^{i}$ and it is different from the fifth equation (6.8).

\section{\protect\bigskip DISCUSSION}

\bigskip In this third part of the paper it has been shown that from the
expressions (2.1) a system of first - order nonlinear differential equations
is obtained, for which always a solution $X^{1}=X^{1}(z)$, $X^{1}=X^{1}(z)$, 
$X^{1}=X^{1}(z)$ exists. Thus the dependence on the generalized coordinates $%
X^{1}$\textbf{, }$X^{2}$\textbf{, }$X^{3}$ in the uniformization functions
(2.1 ) dissappears and only the dependence on the complex coordinate $z$
remains, as it should be for uniformization functions.

Moreover, the initial assumption $dX^{i}=0$ for obtaining the solutions
(2.1) allows us to derive a system of nonlinear differential equations also
for the initial variables $x^{1}$, $x^{2}$, $x^{3}$ and thus the
corresponding solutions $x^{1}=x^{1}(z)$, $x^{2}=x^{2}(z)$, $x^{3}=x^{3}(z)$
in principle can be found. This analysis has been performed in section 3. In
fact, it can easily be guessed that if we have the solutions $X^{1}=X^{1}(z)$%
, $X^{2}=X^{2}(z)$, $X^{3}=X^{3}(z)$ and the additional condition $%
d^{2}X^{i}=0$ (which in fact relates the generalized and the initial sets of
coordinates), then the solutions $x^{1}=x^{1}(z)$, $x^{2}=x^{2}(z)$, $%
x^{3}=x^{3}(z)$ should also be \textquotedblright
coordinated\textquotedblright\ with the previous ones. Indeed, this is
evident from the dependence of the functions $\overline{S}_{1}^{(i)}$,$%
\overline{S}_{2}^{(i)}$,....,$\overline{S}_{6}^{(i)}$ in the system (3.11)
for $\frac{dx^{i}}{dz}$ both on the functions $\frac{\partial F_{i}}{%
\partial X^{k}}$ and $\frac{\partial F_{i}}{\partial x^{k}}$ , i.e. on both
system of coordinates. Of particular importance is the conclusion at the end
of Section 3 that the two sets of coordinates $X^{1}$, $X^{2}$, $X^{3}$ and $%
\ x^{1}$, $x^{2}$, $x^{3}$ should not be treated on an equal footing. This
means that if $dX^{1}$, $dX^{2}$, $dX^{3}$ satisfy the originally derived
cubic algebraic equation, then it is not necessary to assume this for $\
dx^{1}$, $dx^{2}$, $dx^{3}$.

Much more interesting is the other investigated case in Sections 4 - 6,
where a pair of complex coordinates $z,v$ has been introduced and thus
through the generalized coordinates $X^{1}=X^{1}(z,v)$, $X^{2}=X^{2}(z,v)$, $%
X^{3}=X^{3}(z,v)$ a complex structure of the metric tensor components is
introduced. For the investigated case under the assumption $d^{2}X^{i}=0$,
there is only one way for introducing this complex structure - namely,
through the dependence of the initial coordinates on $z$ and $v$, i. e. $%
X^{i}=X^{i}(\mathbf{x}(z,v),z)$. Otherwise, if some other possibility is
assumed, for example $X^{i}=X^{i}(\mathbf{x}(z),z,v)$, then, as proved in
section 4, the obtained system of equations is contradictory. Therefore, it
remains to investigate the full system of equations for the only allowed
case $X^{i}=X^{i}(\mathbf{x}(z,v),z)$, which has been performed in Sections
5 and 6. Remarkably, a nonlinear differential equation is obtained only for
the initial coordinates. However, no such an equation only for the
generalized coordinates can be obtained - the derived equation depends in a
complicated manner on both system of coordinates. Since the existence of
these noncontradictory systems of equations confirms that a complex
structure can be introduced, one may express the line element $ds^{2}=g_{ij}(%
\mathbf{X})dX^{i}dX^{j}$ as 
\begin{equation}
ds^{2}=\widetilde{g}_{zz}(z,v)(dz)^{2}+\widetilde{g}_{zv}(z,v)dzdv+%
\widetilde{g}_{vv}(z,v)(dv)^{2}\text{ \ \ \ \ ,}  \tag{7.1}
\end{equation}%
where 
\begin{equation}
\widetilde{g}_{zz}(z,v)\equiv g_{ij}(\mathbf{X}(z,v))\overset{.}{X}^{i}%
\overset{.}{X}^{j}\text{ \ \ ; \ \ }\widetilde{g}_{vv}(z,v)\equiv g_{ij}(%
\mathbf{X}(z,v))X^{^{\prime }i}X^{^{\prime }j}\text{\ \ \ \ ,}  \tag{7.2}
\end{equation}%
\begin{equation}
\widetilde{g}_{zv}(z,v)\equiv g_{ij}(\mathbf{X}(z,v))\left[ \overset{.}{X}%
^{i}X^{^{\prime }j}+X^{^{\prime }i}\overset{.}{X}^{j}\right] \text{ \ \ .} 
\tag{7.3}
\end{equation}%
This result will be of particular importance in reference to possible
physical applications, which will be considered in another paper. For
example, the linear element of a unit surface in the \textit{Lobachevsky
space with a constant negative curvature} $-\frac{1}{R^{2}}$ can be
represented as [9] 
\begin{equation}
ds^{2}=R^{2}\frac{(a^{2}-w^{2})du^{2}+2uwdudw+(a^{2}-u^{2})dw^{2}}{%
(a^{2}-u^{2}-w^{2})^{2}}\text{ \ \ \ ,}  \tag{7.4}
\end{equation}%
which by means of a suitable coordinate transformation can be brought to the
form 
\begin{equation}
ds^{2}=d\rho ^{2}+e^{-\frac{2\rho }{R}}d\sigma ^{2}\text{ \ \ \ \ .} 
\tag{7.5}
\end{equation}%
The metric (7.5) is the standard form of the three - dimensional Lobachevsky
metric [10] ($d\sigma ^{2}$ is a two - dimensional surface element), where
the ratio $\frac{2\rho }{R}$ may or may not be identified with the
Lobachevsky constant $k=\frac{1}{c}$ ($c$ is a natural unit length element
in the Lobachevsky space). This is particularly important to be mentioned in
reference to Randall - Sundrum models and theories with extra dimensions,
which are based on the \textquotedblright
multi-dimensional\textquotedblright\ analogue of the Lobachevsky metric
(7.5). Some physical applications of the algebraic geometry formalism in
these models will be considered in a separate paper.

\section*{\protect\bigskip Acknowledgments}

The author is grateful to Dr. L. K. Alexandrov, St. Mishev and especially to
Prof. V. V. Nesterenko, Dr. O. Santillan (BLTP, JINR, Dubna) and to Prof.
Sawa Manoff (INRNE, BAS, Sofia) for valuable comments, discussions and
critical remarks. \ 

This paper is written in memory of \ Prof. S. S. Manoff (1943 - 27.05.2005)
- a specialist in classical gravitational theory and physics. \ 

The author is grateful also to Dr. A. Zorin (LNP, JINR) and to J. Yanev
(BLTP, JINR) for various helpful advises and to Dr.V. Gvaramadze (SAI, MSU,
Moscow) and his family for their moral support and encouragement.

\end{document}